\documentclass[showpacs,aps,prd,nofootinbib,showkeys,unsortedaddress,twocolumn]{revtex4-1}
\usepackage{epsfig}
\usepackage{graphicx}
\usepackage{amsmath,amssymb,amsfonts}
\usepackage{array}
\usepackage{url}
\usepackage{hyperref}
\usepackage{multirow}
\usepackage{float}
\usepackage{lineno}
\usepackage{xspace}
\usepackage[usenames,dvipsnames]{color}

\newcommand{\bef}{\begin{figure}}
\newcommand{\eef}{\end{figure}}
\newcommand{\bc}{\begin{center}}
\newcommand{\ec}{\end{center}}

\newcommand{\be}{\begin{equation}}
\newcommand{\ee}{\end{equation}}
\newcommand{\bea}{\begin{eqnarray}}
\newcommand{\eea}{\end{eqnarray}}

\usepackage[usenames,dvipsnames]{color}
\usepackage{multirow}
\definecolor{kugray5}{RGB}{224,224,224}
\usepackage{adjustbox}
\usepackage{booktabs}
\usepackage{multirow}
\usepackage{rotating,tabularx}
 \usepackage{lineno}
 
 \usepackage{hyperref}

\begin{document}
\title{Non-Extensive Statistics in Free-Electron Metals and
  Thermal Effective Mass}
\author{A. Khuntia}
\email{Presently at: The H. Niewodniczański Institute of Nuclear Physics, Polish Academy of
Sciences, 31-342 Cracow, Poland}
\author{Gayatri Sahu}
\author{R. Sahoo}
\email{Corresponding Author Email: Raghunath.Sahoo@cern.ch}
\affiliation{Discipline of Physics, School of Basic Sciences, Indian Institute of Technology Indore, Indore- 453552, INDIA}
\author{D. P. Mahapatra}
\email{Email: dpm.iopb@gmail.com}
\author{N. Barik}
\affiliation{Department of Physics, Utkal University, Vani Vihar, Bhubaneswar-751004, INDIA}

\begin{abstract}

We have applied the non-extensive statistical mechanics to free
electrons in several metals to calculate the electronic specific heat
at low temperature. In this case, the Fermi-Dirac (FD) function is 
modified from its Boltzmann-Gibbs (BG) form, with the exponential 
part going to a 
$q$-exponential, in its non-extensive form. In most cases, the 
non-extensive parameter, $q$, is found to be greater
than unity to produce the correct thermal effective mass, $m^*$, of
electrons. The ratio $m^*/m$ is found to show a nice systematic 
dependence on $q$. Results indicate, electrons in metals, in the presence 
of  long range correlations 
are reasonably well described by Tsallis statistics.

 \end{abstract}
\date{\today}
\maketitle
\section{Introduction}
\label{intro}
Study of properties of metals has great importance in condensed matter
physics. Over two third of solids are metals and, they are good for
both electrical and thermal conduction. Even to understand the
non-metals we should understand the behaviour of metals. The Drude
model,  which applies the kinetic theory of gases to electrons in
metals is successful in explaining basic transport properties in metals. 
It basically is based on the {\it independent} as well
as {\it free-electron} approximations 
ignoring the electromagnetic electron-electron and electron-ion 
interactions. Despite the oversimplification of the reality, it describes quite well
the electrical and thermal conductivities of metals and the Hall
effect \cite{Book:Mermin}. However, Drude model fails to explain many
other observations. According to  Drude  model the specific heat 
due to electrons in a metal is
independent of temperature and can be represented as $C_V=3nk_B/2$, 
where $n$ is the electron density, $k_B$ representing the Boltzmann
constant. This shows metals
with more number of free electrons should have much larger heat
capacity. However, this is not experimentally observed and  is the
consequence of taking  Maxwell-Boltzmann distribution to describe the
energy distribution in a many-electrons system. Later, Sommerfeld
applied the same principle to metals but with a modification to the
electronic velocity (or momentum) distribution, which is taken to be a quantum
FD distribution, derived from Boltzmann-Gibbs (BG) statistics
The drude-Sommerfeld model remains valid and also allows us to
understand many phenomena in metals despite the lack of accuracy and
simplicity of its assumptions. But the key feature of this model is that
electrons in a metal, as described by the FD distribution, are assumed 
to be in thermal equilibrium with the surroundings by collisions. 

However, $C_V$, the specific heat due to the electrons, calculated
using the
above Drude-Sommerfeld model of free electrons is known to  deviate
from experimental data. In text books this deviation is usually
explained in terms of a {\it thermal effective mass} of the electrons.
In an actual situation, there are interactions and collisions
  (electron-electron, electron-ion etc) with long range correlations.
In such a case a non-extensive (NE) Tsallis statistics  
\cite{Book:Tsallis}, instead of BG
statistics, may be of use. 

It must be mentioned here that BG statistics applies correctly to
systems that satisfy broadly the following conditions. The interactions 
must be short ranged, Markovian type, boundary conditions
must be smooth, and there must be no mesoscopic dissipation taking
place. Violation of any or more of these conditions would necessitate
the application NE statistics as proposed by Tsallis.
Tsallis statistics leads to a generalization of the BG entropy, 
which takes care of non-extensivity.
In this formulation, the generalised entropy, is written as

\begin{equation}
S_{q} =k \frac {1-\sum_{i=1}^{W} ~p_{i}^q}{q-1},
\label{eq:Tsallis:entopy}
\end{equation}
where $k$ is a positive number, $p_i$ being the probability of
finding the $i^{th}$ microstate.
Here  $q$, the non-extensive parameter, is 
a real number. In the limit $q\rightarrow$1,  we recover the BG
entropy.  The most
striking part of the Tsallis entropy, $S_q$, is non-additivity, which means
entropy of a mixture of two independent subsystems A and B is not a sum of
entropies of the individual systems, rather 

\begin{equation}
S_{q}^{A +B} = S_{q}^A + S_{q}^B + (1-q )S_{q}^A S_{q}^B,
\label{eq:Tsallis:entopy:2}
\end{equation}
which reduces to BG entropy in the limit q$\rightarrow$1
in which case the additivity nature of entropy is recovered.

Interestingly, Tsallis
statistics finds applications in solid state physics\cite{Ourabah}, 
high energy heavy-ion, as well as
$e^+ + e^-$ collisions. The transverse momentum spectra of the
secondaries created in high energy  $p + p (\bar p)$ 
\cite{Cleymans:2011in,Cleymans:2013rfq,Cleymans:2012ya,Khuntia:2017ite,Azmi:2015xqa}, 
$e^{+} + e^{-}$ collisions \cite{Bediaga:1999hv,Urmossy:2011xk} are
better described by the non-extensive Tsallis statistics\cite{Tsallis:1987eu}. 
In addition, non-extensive statistics with radial flow successfully describes 
the spectra at intermediate $p_T$ in heavy-ion collisions 
\cite{Bhattacharyya:2015hya,Bhattacharyya:2016nmb}. Tsallis distributions 
have also been used to study speed of sound in Hadronic matter 
\cite{Khuntia:2016ikm}. The the non-extensive 
parameter $q$ has been shown to be related to temperature
fluctuations\cite{Wilk:prl,Bhattacharyya:2015nwa}. 

In the present paper we have used this to derive the electronic specific
heat, $C_V$ in some typical free-electron metals to provide an
explanation of the  {\it thermal effective mass} of the electrons 
in terms of the NE parameter {$q$}. 
The list of metals considered includes, Li, Na, Al,
K, Ti, Fe, Co, Ni, Cu, Sr, Ag, Cs, Au, Hg and Pb, with 1, 2, 3 and 4
free electrons in the outer shell. 
It must be added here that such calculations, for free
electrons in several metals, have earlier been carried out using
Tsallis statistics \cite{Ourabah}. 
However, there are some problems
with the results which are in disagreement with experimental data.
In view of this we had decided to carry out a fresh calculation of
the same.

\section{Finite Temperature Free Electron Gas and Specific Heat}
\label{sec:sum}   

At zero temperature, electrons would prefer the lowest energy states
to minimize the energy. But following Pauli principle, they occupy
energy states in pairs with opposite spins. If we have $N$ electrons, at
$T=0$, they occupy the lowest $N/2$ states. The energy of the highest
filled state at absolute zero, being the Fermi Energy, $\epsilon_F$. 

At zero temperature, the occupation probability $f(\epsilon)$, for a given quantum
state with energy $\epsilon~$, is 1 if $ \epsilon ~\leq \epsilon_F$ and
0 if $ \epsilon~ \geq \epsilon_F$,
leading to a step function
\begin{equation}
f(\epsilon) \equiv \left\{
\begin{array}{l l}
1&~~\mathrm{if}~~~\epsilon \leq \epsilon_F \\
0&~~\mathrm{if}~~~\epsilon \geq \epsilon_F \\
\end{array} \right.
\label{eq:2}
\end{equation}

At finite temperature interaction between electrons can be understood
in terms of scattering leading to change in momentum or energy. Since
all states with energy below $\epsilon_F$ are occupied scattering
to low energy states are prohibited leading to a fraction of electrons
with energy near $\epsilon_F$ getting excited to states above
$\epsilon_F$.  
\begin{equation}
0< \it{f}(\epsilon) <1,~\mathrm{ at ~finite ~T.}
\end{equation}

 
 The total energy of a system within the free electron model is given by 
 \begin{equation}
 {\it U(T)}=\int_0^ \infty ~\epsilon~\it{f(\epsilon)}~\it{D}(\epsilon) d\epsilon
 \end{equation}
 where 
 \begin{equation}
  \it{f(\epsilon)} =\frac {1}{1+\exp(\frac{\epsilon-\mu}{k_BT})}
 \end{equation}
 is the FD distribution $f(\epsilon)$ for non-interacting electrons
at  finite temperature. 
Here $\mu$ and $D(\epsilon)$ are the chemical potential and density of
states respectively. Since electrons near the Fermi energy take
the extra energy, $D(\epsilon)$ is to be evaluated at $\epsilon~=\epsilon_F$.
 
The heat  capacity $C_V$, due to the electrons can be calculated using
the following   
\begin{equation}
\it C_V= \frac {\partial U}{\partial T} |_V= -\int_0^ \infty ~\frac{(\epsilon-\mu)^2}{T}~\frac {d\it{f(\epsilon)}}{d\epsilon}\it{D}(\epsilon) d\epsilon
\end{equation}
where $D(\epsilon)$ 
is given by,
\begin{equation}   
\it{D}(\epsilon) =\frac{dN}{d\epsilon}= \frac{3N}{2\epsilon_F} .
\end{equation}

The specific heat can be expressed as 
\begin{equation}   
C_V = \frac{\pi^2}{2} ~ N~k_B \frac {k_B~T}{\epsilon_F}
\end{equation}

\bef[ht]

\includegraphics[scale=0.45]{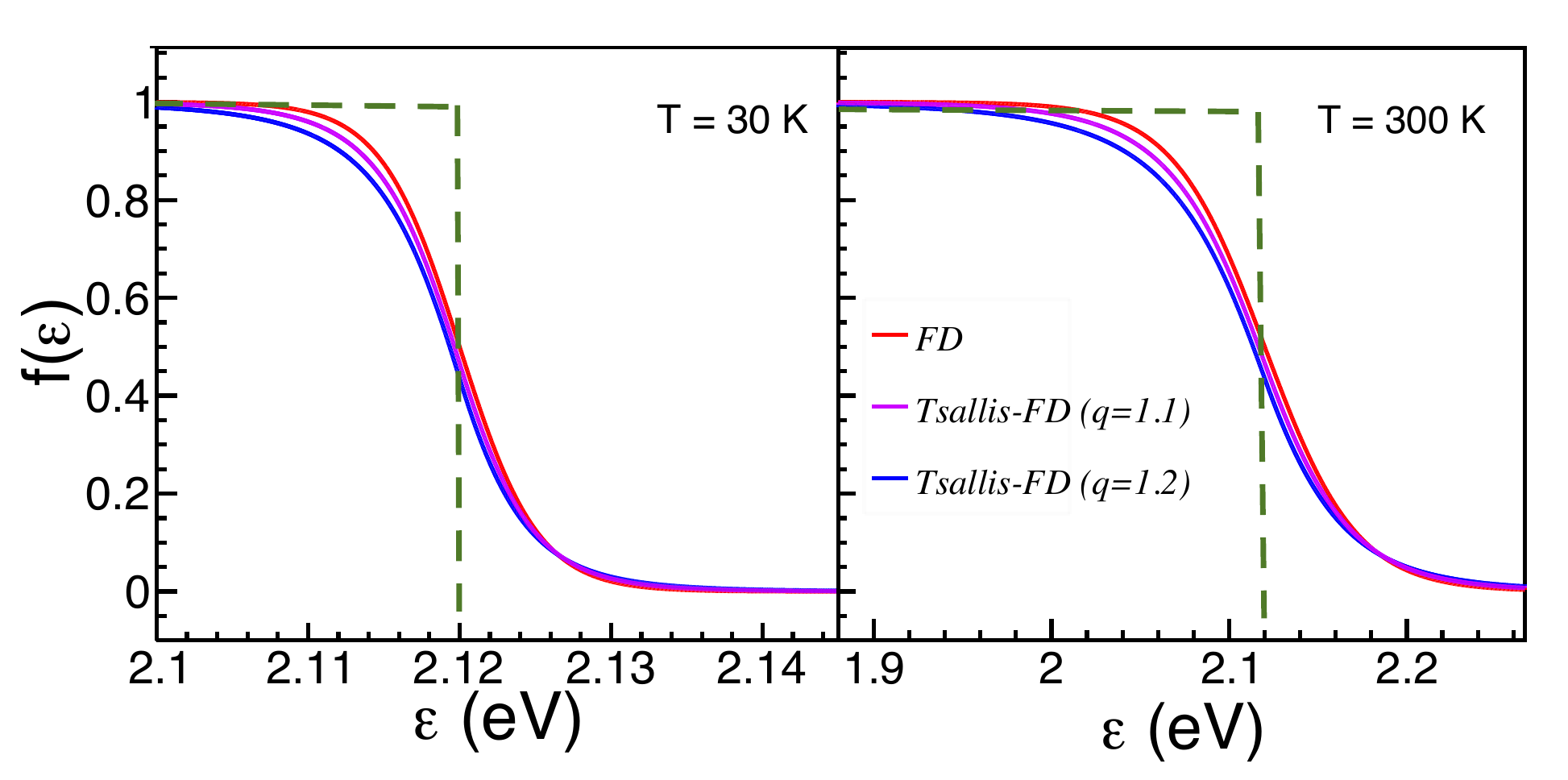}

\caption  {(Color online) Tsallis FD distribution of electrons in potassium
   with $\epsilon_F= 2.12~ eV$, for $q$=1.1 and 1.2 at two different
   temperatures, 30 K and 300 K. The dotted lines show the standard FD distribution
for BG statistics at $T=0$ for comparison. For clarity all distributions
are shown near the Fermi energy.}

 \label{fig:FD:BGTS}  
 \eef

\section{Tsallis Non-Extensive Statistics and Specific Heat}
\label{Tsallis}
In the NE thermodynamics following Tsallis formalism, one uses a $q$-deformed
quantum distribution using the $q$-exponential function (given a step
later). The nonextensivity enters through the parameter $q$, usually
close to unity. The relevant distribution in the present case is a modified FD 
distribution \cite{Cleymans:2011in} as given by

\begin{eqnarray}
\label{FDq}
\it{f_T(\epsilon)}= \frac{1}{1+exp_q(\frac{\epsilon-\mu}{k_{B}T})},
\end{eqnarray}
where the {\it q-exponential} function, $\exp_q(x)$, is defined as
\begin{equation}
\exp_q(x) \equiv \left\{
\begin{array}{l l}
\left[1+(q-1)x\right]^{1/(q-1)}&~~\mathrm{if}~~~x > 0 \\
\left[1+(1-q)x\right]^{1/(1-q)}&~~\mathrm{if}~~~x \leq 0 \\
\end{array} \right.
\label{tsallis-fd1}
\end{equation}

In the limit $q \rightarrow 1$, the q-exponential distribution reduces to the standard exponential form
$$
\lim_{q\rightarrow 1}\exp_q(x)\rightarrow \exp(x).
$$

Hence, the Tsallis NE distribution function in the limit of the NE
parameter, $q \rightarrow 1$ goes over to the standard equilibrium FD
distribution function. The parameter $q$ is a measure of
the degree of non-extensivity based on the amount of correlations and
possible temperature fluctuations that is not included in a pure free electron
picture as described by the FD distribution.
As has been mentioned earlier, in an actual case there are long range
electron-ion and electron-electron correlations. 
The use of NE statistics will be handy for such systems,
where the Tsallis form of FD distribution as given in 
Eq.~\eqref{FDq} could be used.  

However, as shown elsewhere,\cite{Cleymans:2012ya,Conroy} 
just using the $q$-deformed FD
distribution function is not enough to guarantee thermodynamic
consistency. In other words, the formalism must comply with
standard relationships between relevant thermodynamical variables
such as entropy, temperature, energy etc. This requires two additional
constraints $\Sigma{{n_i}^q} = N$ and $\Sigma{{n_i}^q\epsilon_i} = E$
where $n_i$ and $\epsilon_i$ represent the mean occupation
probability of the $i^{th}$ quantum state with energy $\epsilon_i$. $E$
and $N$ represent the total number of particles and the total
energy respectively.

\begin{table*}[tp]

\newcommand{\tabincell}
\centering
\caption{Present list of metals, with atomic number, electron
  density, $\gamma$ and the estimated $q$ values. Numbers in the
  parentheses indicate references.}
\label{table:parameters}

\begin{tabular}{|c|c|c|c|c|c|c|c|c|c|c|}
\toprule

\hline

\multicolumn{1}{|c|}{$\bf{Elements}$} & {$~z~~$} &Electron-density (10$^{28}$/m$^3$)&$\epsilon_{f} $ (eV)& {$ \gamma_{0} ~[\rm{mJ/(mol~ K^2)]}$}&{$ {\gamma_q}~[\rm{mJ/(mol~ K^2)]}$} & {$\gamma_{Exp}~[\rm{mJ/(mol~ K^2)]}$}& { $m^*/m_0$} & {$~q~$} 
\rule{0pt}{3ex}\\
\hline

\rule{0pt}{3ex} 
\multirow{1}{*}{Li}&3&4.70&4.74&0.746&1.649&1.65 \cite{metals:cv}&2.211&1.231\\

\hline
\rule{0pt}{3ex}
\multirow{1}{*}{Na}&~11~&2.65&3.24&1.091&1.381&1.38 \cite{metals:cv}&1.265&1.082\\

\hline
\rule{0pt}{3ex}
\multirow{1}{*}{Al}&~13~&18.1 &11.7&0.906&1.352&1.35 \cite{metals:cv}&1.492&1.132\\

\hline
\rule{0pt}{3ex}
\multirow{1}{*}{K}&~19~&1.40 &2.12&1.667&2.085&2.08 \cite{metals:cv}&1.250&1.078\\
                   
\hline
\rule{0pt}{3ex}
\multirow{1}{*}{Ti}&~22~&9.80&7.75&0.912&3.361&3.36 \cite{metals:cv}&3.684&1.340\\

\hline
 \rule{0pt}{3ex}
\multirow{1}{*}{Fe}&~26~&17.0&11.17&0.633&4.893&4.90 \cite{metals:cv}&7.729&1.505\\

\hline
 \rule{0pt}{3ex}
\multirow{1}{*}{Co}&~27~&15.63&10.58&0.668&4.410&4.40 \cite{metals:cv}&6.599&1.467\\

\hline
 \rule{0pt}{3ex}
\multirow{1}{*}{Ni}&~28~&18.2&11.74&0.602&7.032&7.04 \cite{metals:cv}&11.676&1.617\\

\hline
 \rule{0pt}{3ex}
\multirow{1}{*}{Cu}&~29~&8.47&7.0&0.505&0.690&0.69 \cite{metals:cv}&1.366&1.106\\

\hline 
\rule{0pt}{3ex}
\multirow{1}{*}{Sr}&~38~&3.55&3.95&1.790&3.645&3.64 \cite{metals:cv}&2.036&1.212\\

\hline
\rule{0pt}{3ex}
\multirow{1}{*}{Ag}&~47~&5.86&5.49&0.644&0.648&0.64 \cite{metals:cv}&1.006&1.0002\\

\hline
\rule{0pt}{3ex}
\multirow{1}{*}{Cs}&~55~&0.91&1.59&2.223&3.204& 3.20 \cite{WLien}&1.441&1.122 \\

\hline 
\rule{0pt}{3ex}
\multirow{1}{*}{Au}&~79~&5.90&5.53&0.639&0.692&0.69 \cite{metals:cv}&1.083&1.028\\

\hline 
 \rule{0pt}{3ex}
\multirow{1}{*}{Hg}&~80~&8.65&7.13&0.992&1.865&1.86  \cite{metals:cv}&1.880&1.193\\

\hline 
\rule{0pt}{3ex}
\multirow{1}{*}{Pb}&~82~&13.2&9.37&1.509&2.996&2.99\cite{metals:cv}&1.985&1.206\\

\hline

 \end{tabular}

 \end{table*}

Using Tsallis NE statistics with the above mentioned constraints 
the total energy is given by,
\begin{equation}
{\it U(T)}=\int_0^ \infty ~\epsilon~\it{f_T^q(\epsilon)}~\it{D}(\epsilon) d\epsilon
\end{equation}
where the $q^{th}$ power of the occupation number, $f_T(\epsilon)$, is
is used.
\begin{eqnarray}
\it{f_T^q(\epsilon)}= \Bigg(\frac{1}{1+exp_q(\frac{\epsilon-\mu}{T})}\Bigg)^q.
\end{eqnarray}
So, the heat  capacity is 
 
\begin{equation}
\label{CV:Tsallis}
\it C_V= \frac {\partial U}{\partial T} |_V=- \int_0^ \infty
~\frac{(\epsilon-\mu)^2}{T}~\frac
{d\it{f_T^q(\epsilon)}}{d\epsilon}\it{D}(\epsilon) d\epsilon
\end{equation}
where $D(\epsilon)$ is the density of states at  $\epsilon_F$, given
earlier in Eq.~( 9).
$C_V$ due to electrons can now be calculated from
 eqn. \ref{CV:Tsallis} using
numerical techniques as used in the present case. 

In Fig. \ref{fig:FD:BGTS}, we show the FD distributions as a 
function of energy  for both BG and Tsallis statistics,
respectively for the case of potassium which is a nearly perfect
free electron metal with an $\epsilon_F$ value of $2.12~ eV$. Results 
shown for the Tsallis statistics correspond to two different
temperatures 30 K and 300 K. These results shown
have been 
computed for two different $q$ values {\it viz} 1.1 and 1.2. For
clarity we show only results near the Fermi energy, $\epsilon_F$
 where differences are very clearly seen. The corresponding result 
for the standard FD distribution coming from BG statistics, at $T=0$,
 is also shown for comparison. 
One can clearly notice, at a given temperature, non-extensivity
results in exciting more electrons to higher energy states, as compared 
to BG statistics based FD distribution. The net effect of this is an
increased contribution to $C_V$. This effect becomes more and 
more important with increase in $q$. 

\section{Experimental Heat capacity in a real metal}
In a real metal, at low temperature the total heat capacity can be more precisely written as
\begin{equation}
C_V=\gamma~ T+ A~T^3
\end{equation}
where $\gamma$ and $A$ are constants and related to the material. 
Here, the first linear term is the electronic contribution in which we
are interested, the second cubic term coming from lattice vibrations, 
based on the Debye model (phonons contribution). Therefore, at
low temperatures, 
$C_V/T$ when plotted against $T^2$ results in a straight line. In such a case
the zero intercept corresponding to $T=0$ yields the parameter $\gamma$.
A compilation of this on a large number of systems is available 
\cite{metals:cv,WLien}.

For a free electron gas, that obeys the FD distribution, we
rename the coefficient $\gamma$ as $\gamma_0$. This is given by
\begin{equation}   
\gamma_0 = \frac{\pi^2}{2} ~ \frac {N~ k_B^2}{\epsilon_F}
\end{equation}
where 
\begin{equation} 
\epsilon_F=\frac{\hbar^2}{2m}~k_{F}^2= \frac{\hbar^2}{2m}\Bigg(3~\pi^2\frac{N}{V}\Bigg)^{2/3}.
\end{equation} 
So that
\begin{equation} 
\gamma_0=m\frac{\pi^2~N k_B^2 }{\hbar^2\Bigg(3~\pi^2\frac{N}{V}\Bigg)^{2/3}} 
\end{equation} 

One can now see, the parameter $\gamma_0$ is proportional to the
free electron mass. Following the same prescription
as valid for free electrons obeying the FD distribution, we
can now determine $\gamma$ for electrons following the Tsallis modified 
FD distribution as given in Eq.~\eqref{FDq}. We call it
$\gamma_q$. This is possible
provided one knows the value of the parameter $q$. In such a case the
ratio $\gamma_q/\gamma_0$ can be seen as the ratio of the {\it thermal
  effective mass} 
of the electron, $m^*$, to free electron mass, $m$. 
In other words for a given system obeying the NE statistics, the
parameter $q$ can be determined from a knowledge of the $m^*/m$ 
mass ratio. This is what has been followed in the present work.

\bef[ht]
 \includegraphics[scale=0.4]{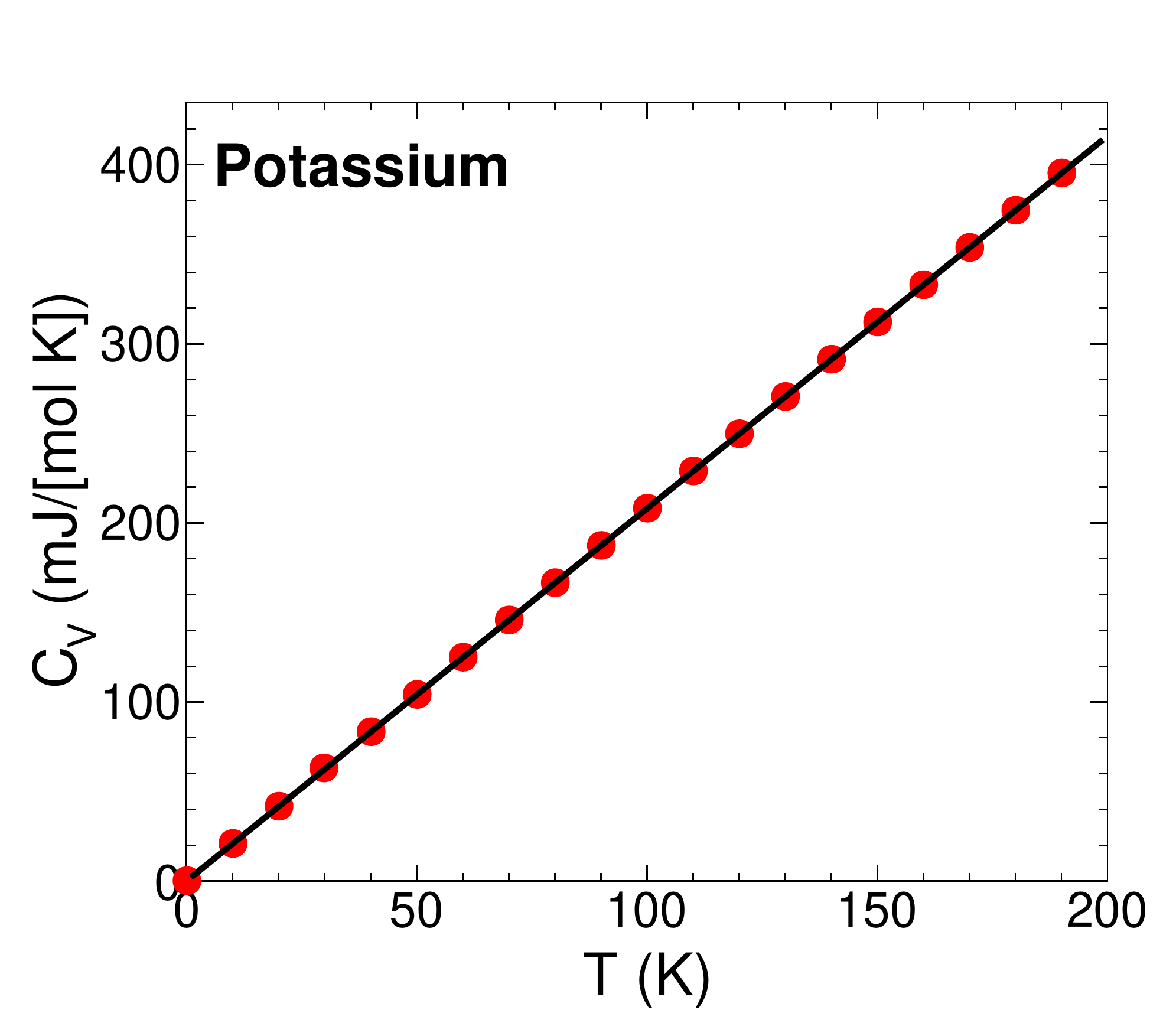}
\caption{Electronic $C_V$ versus $T$ for potassium for $q$=1.078. The slope leads to
  a $\gamma_q$ value of 2.085$~\rm {mJ/(mol.K^2)}$ close to the
  experimental data.}
 \label{fig:cvbyt}  
 \eef

\bef[ht]
 \includegraphics[scale=0.45]{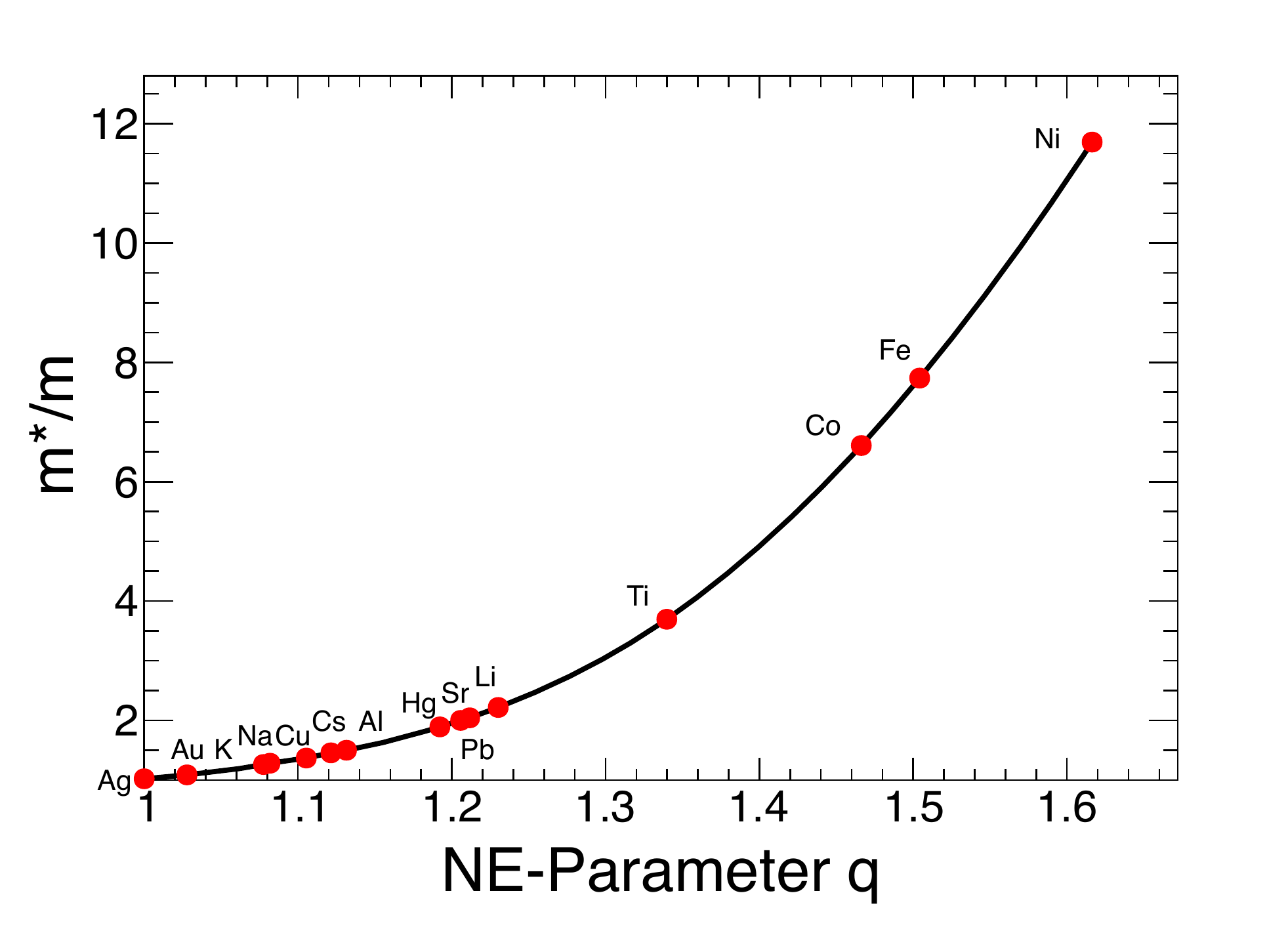}
\caption{Ratio $m^*/m$ for various metals with different $q$ values.
The data are well described by the 
the continuous line obtained using the reduced energy $\epsilon/\epsilon_F$ as 
mentioned in the text. }
\label{fig:mstbym}  
 \eef

 \section{Results and Discussions}

 In our analysis we first look at the case with potassium, a nearly 
perfect free electron system. In this case the total $C_V$ including 
the lattice contribution (as given by Eq. (16)) has been obtained
experimentally as a function of temperature, $T$ \cite{WLien}. 
In the low $T$ region, $C_V/T$ vs $T^2$ is shown to be perfectly linear. 
At $T~=~0$,
$C_V/T=\gamma$ corresponds to the intercept which has a value of
$2.08~ \rm{mJ/(mol.K^2)}$. Taking $\epsilon_F = 2.12~ eV$ as valid 
for potassium we
estimate the parameter $q$ to have a value of 1.078 which is greater
than 1. In this estimation, varying the parameter $q$, we first
determine $C_V$ for a set of temperatures. Since there is no lattice
contribution $C_V$ vs $T$, in accordance with Eq. (15), is perfectly 
linear with a zero intercept, the slope leading to $\gamma_q$. 
Our estimated $C_V$ data for potassium, for $q$ value of 1.078,
 at various temperatures is shown in Fig. \ref{fig:cvbyt}. 
It has a slope ($\gamma_q$) value of 2.085, very close 
to the experimental data. 

The same procedure is followed for the estimation of $q$ and
$\gamma_q$ for other metals using the available data. As mentioned
earlier, our full list 
includes Li, Na, Al, K, Ti, Fe, Co, Ni, Cu, Sr, Ag, Cs, Au, Hg and Pb, 
with 1, 2, 3 and 4 free electron systems as a complete set.
The values, together with, other 
physical parameters such as
electronic density, $\epsilon_F$, $\gamma_0$, $\gamma_q$ and the
ratio $\gamma_q/\gamma_0$ ($=m^*/m$) are presented in 
Table \ref{table:parameters}. In the same table $m^*/m$ data and 
the experimental $\gamma_{Expt}$ values available in the literature 
are also included for comparison. One can see, it is possible to
adjust the NE parameter $q$ to obtain $\gamma_q$ values close
to the experimental data.

From the results included in the table, one can further see, for all metals
considered in this paper, with the exception of Ag, the non-extensive
parameter, $q$ deviates from unity, always being higher. This phenomenological parameter
$q$ takes care of long range correlations, resulting in a higher value
of $\gamma_q~(=~C_V/T)$, which in turn results in a higher value of 
effective mass. As has been mentioned earlier, the parameter
  $\gamma_q$ has
been calculated in an earlier work \cite{Ourabah}, for a number
of metals, that include the present ones
as well. The authors have used Tsallis modified FD statistics with
a distribution function $f_T(\epsilon)$, different from what is used
here. 
Their obtained $\gamma_q$ values for various $q$  (shown 
in Table 1 of ref \cite{Ourabah}) are different from what we get 
from the present calculations. Further, there is no agreement
with experimental data \cite{metals:cv}.
Even the values obtained for $q~=1$, corresponding to pure FD
distribution, are in total disagreement 
with free electron data, some of which are given in the present 
paper as $\gamma_0$. This is not the case with our calculations.

At the end, in  Fig.\ref{fig:mstbym} we show a plot of the ratio 
$m^*/m$ as a function of the NE parameter $q$ as obtained in 
various metals considered here. The results show a nice systematic 
dependence, all data corresponding to different values of $\epsilon_F$
indicating a smooth functional dependence. 
The parameter $q$ distinguishes one system from another as does 
$\epsilon_F$. To check this further, for various values of $q$ in
the range 1-1.7,
 we repeated the calculations for each of the metals considered, 
changing the integration variable to 
$\epsilon/\epsilon_F$ in the expression for $C_V$. In all the cases
the scaled distribution functions fall off to zero near 
$\epsilon/\epsilon_F=1$. This actually removes the dependence
on $\epsilon_F$ bringing out the $q$ dependence. The $\gamma_q$ 
values obtained for each system were then divided by the corresponding 
$\gamma_0$ value leading to the $m^*/m$ ratio as a function of $q$. 
Such a calculation, for each of the systems considered, results in the 
same smooth curve shown, passing exactly through all the points.
 This means, $m^*/m$ for all 
free-electron-metals, irrespective of their Fermi energies, follow 
a universal trend in terms of the parameter $q$. 

\section{Summary}

In this paper, the deviation of specific heat in the free electron
model from the experimentally observed value in a number of 
metals is explained 
using the Tsallis form of FD distribution function.
 The phenomenological NE parameter $q$, in 
the $q-$deformed FD distribution has been fitted to produce 
the experimental data.
With the exception of Ag, in all
cases considered, the non-extensive parameter, $q$, has been found to
be greater than unity.
This is mainly due to long range correlations
present in the electron system.
We have also obtained a nice systematic dependence of
$m^*/m$ as a function of the NE parameter $q$, a behaviour similar
to which, has not been reported.

\section{Acknowledgement}
GS would like to acknowledge the financial supports from the Department of Science and Technology, Govt. of India
through its Women Scientist Scheme (Project No: SR/WOS-A/PM-1/2017). AK and RS acknowledge the financial supports from ALICE Project No. SR/MF/PS-01/2014-IITI(G) of Department of Science \& Technology, Government of India.


\begin{thebibliography}{}

\bibitem{Book:Mermin} Solid State Physics, N. W. Ashcroft and N. D. Mermin, Cengage Learning (2003).

\bibitem{Book:Tsallis} Introduction to Nonextensive Statistical Mechanics: Approaching a Complex World, C. Tsallis, Springer Publishing (2009).

\bibitem {Ourabah}  
K~Ourabah and M~Tribeche,
International Journal of Modern Physics B {\bf 27}, 1350181 (2013).

\bibitem{Khuntia:2017ite} 
  A.~Khuntia, S.~Tripathy, R.~Sahoo and J.~Cleymans,
  Eur.\ Phys.\ J.\ A {\bf 53}, 103 (2017)
  
\bibitem{Azmi:2015xqa} 
  M.~D.~Azmi and J.~Cleymans,
  Eur.\ Phys.\ J.\ C {\bf 75}, 430 (2015)

\bibitem{Cleymans:2013rfq} 
  J.~Cleymans, G.~I.~Lykasov, A.~S.~Parvan, A.~S.~Sorin, O.~V.~Teryaev and D.~Worku,
  Phys.\ Lett.\ B {\bf 723}, 351 (2013)

\bibitem{Cleymans:2012ya} 
  J.~Cleymans and D.~Worku,
  Eur.\ Phys.\ J.\ A {\bf 48}, 160 (2012)

\bibitem{Cleymans:2011in} 
J.~Cleymans and D.~Worku,
J.\ Phys.\ G {\bf 39}, 025006 (2012)
  
\bibitem{Bediaga:1999hv} I.~Bediaga, E.~M.~F.~Curado and J.~M.~de Miranda,
 Physica A {\bf 286}, 156 (2000).
 
 \bibitem{Urmossy:2011xk} K.~Urmossy, G.~G.~Barnafoldi and T.~S.~Biro,
  Phys.\ Lett.\ B {\bf 701}, 111 (2011).
  
 \bibitem{Tsallis:1987eu}  C.~Tsallis,
  J.\ Statist.\ Phys.\  {\bf 52}, 479 (1988).
  
\bibitem{Bhattacharyya:2016nmb} 
  T.~Bhattacharyya, A.~Khuntia, P.~Sahoo, P.~Garg, P.~Pareek, R.~Sahoo and J.~Cleymans,
  Acta Phys.\ Polon.\ Supp.\  {\bf 9}, 177 (2016)

\bibitem{Bhattacharyya:2015hya} 
  T.~Bhattacharyya, J.~Cleymans, A.~Khuntia, P.~Pareek and R.~Sahoo,
  Eur.\ Phys.\ J.\ A {\bf 52},  30 (2016)
  
\bibitem{Khuntia:2016ikm} 
  A.~Khuntia, P.~Sahoo, P.~Garg, R.~Sahoo and J.~Cleymans,
  Eur.\ Phys.\ J.\ A {\bf 52}, 292 (2016)
  
\bibitem{Wilk:prl}
G.~ Wilk, and Z.~ Wlodarczyk, Phys. Rev. Lett. {\bf 84}, 2770 (2000).

\bibitem{Bhattacharyya:2015nwa} 
  T.~Bhattacharyya, P.~Garg, R.~Sahoo and P.~Samantray,
  Eur.\ Phys.\ J.\ A {\bf 52}, 283 (2016)
  
\bibitem{Conroy}
J.~M.~Conroy, H.~G.~Miller, and A.~R.~Plastino,
Phys. Lett A {\bf 374} 4581 (2010).


\bibitem{metals:cv}   
\url{http://www.knowledgedoor.com/2/elements_handbook/electronic_heat_capacity_coefficient.html}
 
 \bibitem{WLien}   
W~H~Lien and N~E~Phillips, Phy. Rev. {\bf 133}, A1370  (1964).   




\end{thebibliography}
\end{document}